\documentclass[12pt]{article}
\usepackage{pdfpages}
\pdfoutput=1
\def\siml{\underline{\sim}}

\def\np{\newpage}
\def\part{\partial}

\def\phih{\hat{\phi}}
\def\sgh{\hat{\sigma}}

\def\jb{\bar{j}}

\def\ybr{\bar{y}}

\def\Ebf{{\bf{E}}}

\def\Sbr{\bar{S}}

\def\ktl{\tilde{k}}

\def\kv{\vec{k}}

\def\th{\theta}

\begin{document}
\begin{center}
{\bf SPECTRAL ESTIMATION OF PLASMA 
FLUCTUATIONS II:\\ 
NONSTATIONARY  ANALYSIS OF ELM SPECTRA} \\
\end{center}
\begin{center}
{\bf Kurt S. Riedel$^*$, Alexander Sidorenko$^*$, \\
Norton Bretz$^{\dag}$, and David J. Thomson$^+$\\
\  \\
 $*$ Courant Institute of Mathematical Sciences,\\ 
New York University,\ New York, New York 10012-1185 \\
$\dag$ Princeton Plasma Physics Laboratory, \ Princeton, NJ 08544-0451 \\
%\medskip
 + AT\&T Bell Laboratories, Murray Hill, NJ 07974-0636  \\
}\end{center}

\begin{abstract}
Several analysis methods for nonstationary fluctuations are described
and applied  to the edge localized mode (ELM) instabilities
of limiter H-mode plasmas. The microwave 
scattering diagnostic observes poloidal $k_{\theta}$ values of 
3.3 cm$^{-1}$, averaged over a 20 cm region at  the plasma edge.
A short autoregressive filter 
enhances the nonstationary component of the plasma fluctuations by 
removing much of the background level of stationary fluctuations.
Between ELMs, the spectrum predominantly consists of broad-banded 
300-700 kHz fluctuations propagating in the  
electron diamagnetic drift direction,
indicating the presence of a negative electric field near the plasma edge.
%is asymmetric, with a local maximum  
The time-frequency spectrogram is computed with the multiple taper technique.
By using the singular value decomposition of the spectrogram, 
it is shown that the spectrum during the ELM is broader 
and more symmetric than 
that of the stationary spectrum.
The ELM period and the evolution of the spectrum between ELMs varies from
discharge to discharge. For the discharge under consideration which
has distinct ELMs with a  1 msec period, 
%as the spectrum relaxes from the broad symmetric spectrum of the ELM.
%We estimate the relaxation time for the return of the
the spectrum has a maximum in the electron drift direction which 
relaxes to a near constant value %its characteristic shape 
in the first half millisecond after the end of the ELM and then grows slowly. 
In contrast, the level of the fluctuations in the
ion drift direction increases exponentially by a factor of eight in the five
milliseconds~after the ELM.
%The ELM onset appears correlated with the ``ion'' spectrum reaching
%a critical level.
High frequency precursors are found which occur one millisecond before 
the ELMs and propagate in the 
ion drift direction. %250 - 650 kHz range.
These precursors are very short ($\sim 10 \mu$secs),
coherent bursts, and they predict the occurrence of an ELM with 
a high success rate.
A second detector, measuring fluctuations 20 cm from the plasma edge with
$k_{\theta}$ values of 8.5 cm$^{-1}$, shows no precursor activity.
The spectra in the ion drift direction are very similar on both detectors,
while the ``electron'' spectrum level is %roughly three times 
significantly larger on this second detector.

\end{abstract}

\ \\

 
\newpage

\noindent
{\bf I. Introduction}
\ \\

Edge localized modes (ELMs) occur in the H-mode phase$^{1,2}$ 
of a tokamak discharge
and are usually associated with periodic loss of confinement 
and transient deposition of heat on the divertor/limiter. 
A variety of different types of ELMs have been observed (and
ELM precursors)$^{1-7}$, and there remains no widespread consensus 
on the nature and causes of ELMs.
Therefore, it is desirable to understand
the physical characteristics of ELMs and if possible to predict their
occurrence. 

In one class of ELM discharges$^{1,5,6}$, a
magnetohydrodynamic (MHD) precursor is seen %for a fraction of a millisec~
prior to the ELM. When the precursor lasts for
many sampling times and consists of a single mode, straightforward analysis
is adequate to resolve the basic features of the ELM precursor. 
Another class of precursors$^{2-5}$, consisting of a transient burst of high
frequency fluctuations, %possibly incoherent precursors, 
has also been observed.
For these short lived bursts, the choice of analysis methods can noticeably
enhance the resolution of the transient events.

In this article, we describe a number of advanced signal processing
techniques and apply these methods to a limiter H-mode discharge
from the tokamak fusion test reactor (TFTR)$^{2,7,8}$. 
Our goal is
to show how these methods give high resolution estimates of transient
plasma fluctuations. In a previous article$^9$, we describe advanced spectral
analysis methods which give highly accurate spectral estimates for
relative short time series. For the transient ELM precursors, the typical
data length is 100 points so that high resolution estimates are
particularly important.

We examine the  evolution of the spectrum between ELMs. The spectral density
fluctuates strongly on the millisecond time scale. This indicates
that there are only a very  small number of waves present in a given 
frequency range at one time. When the spectrum is averaged over a number of
milliseconds,  the fluctuations can be treated as a stationary process.

The time period between ELMs and the spectral evolution between ELMs varies 
appreciably between discharges.
We consider a particular TFTR discharge, \# 49035,
which has particularly well defined ELMs. We caution that other TFTR 
discharges sometimes have distinctly different behavior and that  
%we have not found any systematic 
no single pattern can describe the ELM behavior in all discharges.
%spectral evolution betweenELMs.
%In some discharges, the ``electron'' spectrum grows while the ``ion'' spectrum
%remains constant. 

Our results confirm the findings of Kaye et al.$^3$ and McGuire et al.$^4$ 
that one or
more short bursts of high frequency fluctuations precede the ELM by a
fraction of a millisecond. Our findings give a more precise characterization
of the TFTR ELM precursors:
the precursor bursts occur in the 500 kHz range propagating 
in the ion drift direction, and therefore have a spectral density which is
%We quantify the spectral range of the precursors, and show that it is 
distinctly different from both the stationary spectrum between ELMs 
and the ELM spectrum itself. Our sampling rate is ten times faster than
that in Ref. 3; thus it is understandable that we observe 
precursors with one tenth the duration as Kaye et al$^3$.
We often use the terms ``electron'' spectrum and 
``ion'' spectrum
to denote the poloidal direction of propagation %of the fluctuations 
in the electron and ion diamagnetic drift directions respectively. 

After the ELM, the intensity of the spectrum in the electron drift direction 
grows by 50--80\% during the first half millisecond after the end 
of an ELM.  After the half millisecond, the intensity of the electron
drift spectrum saturates at a more or less  constant value.
%relaxes to a constant value in 0.5-1.0 msec. 
In contrast, {\it the intensity of the spectrum in the ion drift 
direction increases exponentially between ELMs. During the 5-6 milliseconds
between ELMs, this ``ion'' spectrum grows by a factor of eight.} When this
``ion'' spectrum grows more slowly, the onset of the ELM is delayed. 
Thus, in this particular discharge, the ELM onset appears correlated 
with the ``ion'' spectrum reaching a critical level.

In Section II, we describe the TFTR H-mode microwave scattering 
data set$^{10-12}$
which we use as an example. In Section III, we review autoregressive
filters$^{13,14}$ and show how these filters highlight 
nonstationary phenomena. We
consider time-frequency representations of evolutionary spectra$^{15,16}$ 
in the next three sections. 
In Section IV, we compute the singular value decomposition
of the time-frequency distribution to isolate the spectrum during the ELMs.
In Section V, we examine  the evolution of  the spectral density between
ELMs. In Section VI, we use a high resolution, prewhitened, multitapered
evolutionary spectral estimate to examine high frequency precursors to
the ELMs.
%The multitaper SVD technique was previously employed in ---.

\ \\

%\newpage
\noindent
{\bf II. Stationary analysis of the TFTR ELM data set} 
\vspace{.15in}
%\ \\

a) Microwave scattering data set
\vspace{.1in}

%We now describe the TFTR data and the physics of the underlying fluctuations. 
The TFTR microwave transmitter launches a 60 GHz plasma wave 
linearly %elliptically
polarized in the extraordinary mode below the electron cyclotron 
frequency$^{10-11}$ at 112 GHz.
The 60 GHz extraordinary mode plasma wave propagates in the poloidal plane
through the plasma to the edge, where plasma density 
fluctuations scatter the incoming wave$^{}$. 
Figure 1 displays a ray tracing calculation of the
extraordinary mode wave path. The center of the  beam path is the central
curve while the  outer curves  mark the beam $\frac{1}{e}$ power half-width 
of 5.0 cm.
The scattered wave is measured by two detectors %at two heterodyne detectors, 
located near the bottom of the vacuum vessel.
A third detector at the top of the
vessel, known as detector \#1, measures backscattered power
at $|\kv_{scat}| \ \siml\  20\ {\rm cm}^{-1}$, and is not used in this article.

We examine the fluctuation spectrum of TFTR discharge \#49035 as measured by 
the microwave scattering diagnostic$^{10-12}$. 
The detected signal is down-shifted from 60 GHz to 1 MHz  using 
standard intermediate frequency (IF) techniques.
The time series begins 4.2 sec~into the discharge, and
is totally contained in the H-mode phase.
%We redefine time equals zero $(t \equiv 0)$ as the start of our series. 
Our data consist of 524,288 time samples %of the two detectors
with a uniform sampling rate of   5 MHz over the time interval.
Thus, the fluctuations are recorded over a tenth of a second time interval.
The data has been band-pass filtered with an anti-aliasing filter
with a low-pass filter half-width  of 2.5 MHz.

For TFTR discharge  \#49035, the plasma parameters are
the toroidal magnetic field: $B_t = 4.0$ Tesla; 
the edge $q$, $q_a = 5.7$; the total current, $I_p = 0.9$ MA; 
the line average density, $\bar{n}_e = 2.5 \times 10^{13}$ cm$^{-3}$;
and the absorbed power, $P_{NBI}$ = 13 MW. 
The central electron temperature is approximately 6 keV 
and the central ion temperature is approximately 21 keV. 
In the scattering volume, the local plasma parameters are
%... The corresponding physical variables are 
$\rho_i = 0.16$ cm, $\rho_S\ \siml\ \rho_i \sqrt{T_e /T_i} $ = 0.09 cm. 
%$k_{\theta} \siml$ 3.3 cm$^{-1}$, $V_{D_E} = 9.0 \times 10^4 $m/sec, 
%$\omega_{D_e}/2\pi = 57$ kHz. $\omega_{D_i}/2\pi = ?$ kHz.

We concentrate on detector \#3, which measures fluctuations
averaged over a 20 cm region at  the plasma edge.
Figure 1a displays the geometry of the transmitted extraordinary
mode and the receiver antenna pattern.
 The three curves emerging from the bottom 
right are the center of the detector line of sight and its inner and outer
$\frac{1}{e}$ power sensitivity. The ellipses denote the product of the
transmitter power profile and the receiver antenna profile.
%scattering volume of the detector. 
Detector  \#3 measures
$|\kv_{scat}- \kv_{inc}|\ \siml\  3.3\ {\rm cm}^{-1}$ 
and $\kv$ is parallel to the 
poloidal magnetic field at $\frac{r}{a}\ \siml\ 1.0 \pm 0.1$.

Detector \#2 measures fluctuations somewhat farther in the plasma  interior,
at $\frac{r}{a}\ \siml\ 0.75 \pm 0.1$ with 
$|\kv_{scat}- \kv_{inc}|\ \siml\  8.5\ {\rm cm}^{-1}$.
Figure 1b gives the corresponding line of sight information for detector \#2. 
%approximately the middle ? third of the plasma
%while the ELMs are located in the outer third of the plasma.
Unless otherwise specified,  our analysis is based on detector \#3,
and detector \#2 will be used for comparison. Unfortunately,  we  cannot
determine whether the signal differences from detector \#2 to \#3
are due to detector \#2 measuring  smaller wavelength turbulence or 
due to the scattering volume for detector \#2 being located farther in the 
interior of the plasma.
ELM activity is not observed
when the center of the scattering volume is located inside 
$\frac{r}{a}\ \siml\ 0.5$.

These strongly beam-heated discharges are rotating toroidally with a velocity
profile which is peaked on axis with $v_{\phi}(r=0) \sim 10^6 -10^7$ cm/sec.
At the edge, the toroidal rotation is typically very small, 
$v_{\phi} < 3 \times 10^{5}$ cm/sec.  %$f \le 10$ kHz.
Because the fluctuations are primarily aligned perpendicular to %$\vec{k}$,
the total magnetic field, the observed poloidal Doppler shift is 
proportional to
$\Delta \omega\ \siml\ k_{\theta}v_{\phi}(r) {B_{\theta}\over B_{\phi}}$. 
%Thus, we can only describe the plasma spectrum 
%relative to the baseline IF~ frequency and not relative to the 
%absolute frequency in the plasma frame.
Thus, only spectral shifts of less than %of the electron diamagnetic frequency,
$20$ kHz may be due to toroidal fluid motion$^{2}$. %easily observable 
%(at $k_{\perp}\simeq$ 3.3\ cm$^{-1}$).
The ion and electron drift frequencies are approximately 50 kHz
at $\frac{r}{a}\ \siml\ 0.9$ for $\ktl_{\theta}\ \siml\  3.3\ {\rm cm}^{-1}$.
Detailed profile information in the region 
$0.9 \le \frac{r}{a}\ \siml\ 1.0$ is not available, however.
\ \\

%\newpage
{b) Exploratory analysis}
\ \\

We begin by computing the spectrum using the multiple spectral window
analysis as described in Refs.~9, 17-19. We use 
20 orthogonal tapers and then kernel smooth the estimate over
a 20 kHz frequency bandwidth.
In assuming the plasma fluctuations are stationary, we are averaging
the fluctuation spectrum over the short ELM bursts and the long
quasistationary times between ELMs. Figure 2 displays this ``composite''
spectrum which we compute using the smoothed multitaper method$^{9,17,19}$.
The solid curve is the estimated spectrum from detector \#3 and the
dashed curve is from detector \#2. 

On detector \#3,
a frequency shift greater than 1 MHz represents fluctuations which are
traveling poloidally in the electron drift direction in the laboratory frame.
Frequency shifts less than 1 MHz correspond to fluctuations 
moving in the ion drift direction in the laboratory frame.
%In the absence of plasma rotation, the 1 MHz IF~line could be interpreted 
%as the zero frequency, with electron drift fluctuations occurring at higher
%frequencies and ion drift fluctuations occurring at lower frequencies.
Detector \#2 measures  $k_{\theta}$  fluctuations with the $\ktl_{\theta}$
direction reversed in comparison 
with that of detector \#3, and therefore electron drift frequencies
are reversed with respect to 1 MHz on detector \#2. To compensate for this 
reversal of $\ktl_{\theta}\ $
in detector \#2, we have reflected
the frequencies about the 1 MHz line: $ f \rightarrow$ 1 MHz -- $f$.

The spectral peak near 1 MHz is  partially coherent.
%and corresponds to the receiver intermediate frequency. 
%(IF) caused by wall and waveguide reflections.
The broadening of the 1 MHz peak is believed to
be due to intense edge fluctuations$^{9,11,12}$ 
at low $k$, $k_{\perp} \sim 0.1$ cm$^{-1}$. 
The detector sensitivity to these low $k$ fluctuations is  weak,
but the edge fluctuation intensity is sufficiently large that 
broadening is appreciable. The broadening of the 1 MHz line on detector \#3 is 
larger than that of detector \#2 because detector \#3 measures fluctuations
much closer to  the plasma edge.

The spectrum is asymmetric with a larger spectral density on the high
frequency/electron drift side, as is typical of H-mode discharges$^2$. 
%Although there is no secondary  maximum of the fluctuation spectrum, 
There is a plateau in the spectral density from 1300 kHz to 1500 kHz.
A  frequency shift of 300 kHz at $k_{\perp}\simeq$ 3.3 cm$^{-1}$
corresponds to a fluid velocity of 
$\tilde{v}_{\theta} \simeq \frac{2\pi \Delta f}{\ktl_{\theta}} \sim 
6 \times 10^5$ cm/sec. 
The edge toroidal rotation velocity is measured by 
the CHERS diagnostic$^{2,8}$, and is at most 
${B_{\theta}\over B_{\phi}}\times 3\times 10^5$ cm/sec 
$\simeq 3\times 10^4$ cm/sec. 
 Thus, the toroidal rotation is not
responsible for the frequency shift.
The poloidal fluid rotation velocity is not measured, and can account for 
an unknown but significant fraction of the frequency shift. 

The other possible explanation for the asymmetric broadening of the spectrum
is the presence of plasma fluctuations which are propagating 
in the electron drift direction with a frequency range of 300-700 kHz.
These fluctuations have frequencies five to ten times larger than the
electron diamagnetic frequency.
We are unaware of any plasma instabilities with frequencies in this range.
Thus, we believe the most likely cause of the frequency shift is the
presence of a strong poloidal electric field, leading to strong poloidal
fluid motion.
We also measure fluctuations in the ion drift direction in the 500 kHz range. 
One explanation is that the gradient in the poloidal electric field
is sufficiently strong that both the ion and electron frequency shifts are
caused by the poloidal electric field which is 
changing signs within the scattering volume. 
In DIII-D$^{5,20,21}$, 
H-mode electric fields are observed to be negative near the edge
scape-off llayer and to become positive in the interior.

In discharges such as \#49035, we primarily measure the poloidal  component 
of the wave vector. In other discharges where the fluctuations are primarily
measured in the  radial direction, we find the frequency shift is much less.
This supports our belief that the poloidal  electric field is primarily
responsible for the frequency shift. 
%The presence of large shifts in both
%the electron and ion drift directions is explained by a large gradient
%in the poloidal electric field within the scattering volume.

In Figure~2, the spectrum  in the ion drift direction 
{\it decays exponentially in frequency with the same rate in both 
detectors.} To normalize the amplitudes,
we have multiplied the spectrum of detector \#2 by a factor of ten. 
We have chosen this normalization such that 
the amplitudes of the two detectors are the same 
for the frequencies where the spectrum of detector \#3 begins to broaden
rapidly. This normalization is natural because these frequencies constitute
the beginning of the true scattered signal. With this normalization,
the ``ion'' spectrum in both detectors   has the same amplitude
while the electron amplitude is three times larger. Since our normalization
of detector \#2 is somewhat arbitrary, the precise ratio is also
arbitrary. Nevertheless, the electron turbulence is stronger on detector
\#2, and thus the electron turbulence is stronger at
$\frac{r}{a}\ \siml\ 0.75 $ than  
at $\frac{r}{a}\ \siml\ 1.0$, or the electron turbulence is stronger  for 
$\ktl_{\theta}\ \siml\  3.3\ {\rm cm}^{-1}$ %wavelengths of 0.75 cm 
than for $\ktl_{\theta}\ \siml\  8.5\ {\rm cm}^{-1}$. %wavelengths of 2 cm.

Figure 3a displays
the  12 millisecond~data segment from $t = 2$ msec~to $t = 14$ msec.
%of the microwave scattering diagnostic for T.F.T.R. discharge \#49035.
The ELMs occur at $t =$  6.07 and 13.19  msec.
We will show that characteristically,
ELM precursor bursts are centered in the $250-650$ kHz range roughly
1 msec~prior to the ELM.
Two of these precursor bursts in the $250-650$ kHz range occur at
$t =$  5.186 and  12.734 msecs.
But, it is very difficult to identify the $250-650$ kHz 
bursts in the raw data.

The ELMs appear to consist of a series of distinct bursts, and resemble
a sinusoid which is modulated at high frequency. We have tried unsuccessfully
to develop a joint time-frequency representation of the ELM burst.

\ \\

%\newpage
\noindent
{\bf III. Autoregressive filters to enhance nonstationary events}
\ \\

In this section, we assume that the measured time
series, $\{ x_i \}$, is related to an underlying basic time series,
$\{ z_i \}$, through an autoregressive (AR) process$^{13,14}$ of order $p$:
$$
x_t = \sum_{i=1}^p \alpha_i^{(p)} x_{t-i} + z_t 
\ . \eqno (1)$$
%The basic time series, 
$\{ z_i \}$ is called the innovation sequence because
it is usually assumed to be a series independent random perturbations. The 
autoregressive model introduces time correlation into the measured time series
through the autoregressive lag parameters, $\alpha_i^{(p)}$.
We  denote an  autoregressive  filter of length $p$ by AR($p$).

Several different numerical methods for determining the autoregressive lag
parameters, $\alpha_i^{(p)}$, are given in Refs.~13-14. We estimate
the autoregressive parameters  by solving
the Yule-Walker equations, i.e. minimize the residual sum of squares:
$$
\sgh_z^2 = \frac{1}{N-p}\sum_{t=p}^N \left( x_t - \sum_{i=1}^p \alpha_i^{(p)}
x_{t-i} \right)^2  \ .\eqno (2)$$
%To illustrate the stationary character of the drift wave fluctuations, 
The main point of this section is that {\it the autoregressive  filter 
appreciably enhances the nonstationary bursts in the data because
the filter removes the stationary part of the signal.}
Figure~3a displays the raw data  and
Fig.~3b plots the residuals, $\{ z_i \}$, of the autoregressive filter.
%$^{8,10}$,  which nearly removes the 1 MHz peak. Note that 
%Filtered data after using a tenth order autoregressive filter.

In Figure 3b, the ELM amplitude is enhanced relative to the background level 
because the ELM spectrum is broader than the ambient spectrum.
%The intermittent bursts which are concentrated in the 1 MHz range 
%(such as those at  $t =$  2.43 and  3.01)
%are reduced by filtering. In contrast,  
The two precursor bursts in the $250-650$ kHz range, at
 $t =$  5.186 and  12.734 msecs,~have been noticeably enhanced
relative to the raw data of Fig.~3a.
%This indicates that the nonstationarity is concentrated in the 1 MHz  peak.
%At ? millisec~and at ? msec, {\it precursors are apparent in the residual plot.}
The large residuals in Fig.~3b last for roughly 60 data points,
which corresponds to 11 $\mu$sec.
A third precursor in the $250-650$ kHz range,  at 
 $t =$  5.56 msecs,~is less visible in the AR residual plot and
can only be seen in the time-frequency plots of Sec.~VI.

In Figure 3a, there are two large non-ELM bursts, 
at $t =$  2.44 and  3.01 msecs, which 
are  centered in the 1 MHz frequency range. 
Most of the other fluctuation bursts are also due to changes 
in the 1 MHz frequency range (shown in Sec.~VI).
These 1 MHz features are greatly reduced in Fig.~3b due to
the AR filter.

As a second example to illustrate the power of the AR filter to 
emphasize nonstationary bursts, 
we consider the Ohmic sawtooth TFTR discharge \# 50616
which we studied in Ref.~9. For this discharge, the microwave
scattering volume is centered at ${r\over a} = 0.3$ and 
$k_{\theta} =4$ cm$^{-1}$.
%For TFTR discharge number 50616, the plasma parameters are
%$B_t = 4$ Tesla, $I_p = 1.2$MA, $n = 3.6 \times 10^{15}$ cm$^{-3}$. 
%The central electron temperature is approximately 2 keV 
%and the central ion temperature is approximately 1 keV, 
%In the scattering volume, the local plasma parameters are
%$\rho_i = 0.08$ cm, $\rho_S$ = 0.11 cm, $k_{\theta}\ \siml\ 3$ cm$^{-1}$,
%$V_{D_E} = 4.9 \times 10^4 $m/sec, $\omega_{D_e}/2\pi = 23$ KHz.
%The time series begins 5.0 sec~into the discharge, and
%is totally contained in the Ohmic phase.
%We redefine time equals zero $(t \equiv 0)$ as the start of our series. 
%A macroscopic sawtooth oscillation occurs at 
%3.86 msec~into the data set.
%Two macroscopic sawtooth oscillations occur at times $t$ = 0.020, 0.061.
%The scattering volume lies just outside the sawtooth mixing radius.
%Our data consists of 65,500 time samples 
Figure 4a displays the 65,535 point raw data. The sawtooth 
oscillation at 3.86 msec~is barely visible.
The blip  at 9.12  msec~is an unknown and unrelated event.
Fig.~4b plots the residuals of the autoregressive filter %$^{8,10}$
which nearly removes the 1 MHz peak. 
The sawtooth is easily discernable in the filtered data in contrast 
to the raw data.

The variance of the raw data in Fig.~4a 
appears to be increasing in time, thereby
calling into question the assumption of stationarity. 
%To illustrate the stationary character of the broad-banded fluctuations, 
Figure 4b shows that the filtered data has no temporal increase in the
variance.
Thus the increasing variance of the raw data is associated with the
1 MHz IF of the receiver and not with the broad-banded
plasma fluctuations$^{11,12}$.
The changing amplitude of the 1 MHz peak
occurs due to  phase changes between the scattered signal and the local 
oscillator of the detection circuit 
%by vibrations in the vessel wall and the electron drift,
and is not directly related to changes in the plasma fluctuations.
%By using the singular value decomposition of Sec.~IV, we can 
%show that the underlying plasma fluctuations are nearly stationary and 
%are decoupled from the nonstationary behavior of the central peak at 1   MHz. 

The autoregressive model may also be used to detect and correct
statistical outliers. We compare the measured data with
%solated date points which differ from an adequate 
the predicted value given by the AR model. 
When the residual error is many standard deviations large,
we mark/plot the outliers for further scrutiny. 
%we by their predicted value.
In  robust spectral estimation$^{22,23}$, severe statistical outliers are  
replace  by their fitted values. In robustifying the data, we focus  on 
the stationary spectrum and discard the transient burst associated with
the ELMs and precursors. Since we are primarily interested in these transient 
phenomena, we {\it do not robustify} the data.

When both processes are stationary,
the spectra of the measured and innovation processes,
$\{ x \}$ and $\{ z \}$, are related by
$$
S_x (f) = {S_z (f) \over \left| 1 - \sum_{k=1}^p
\alpha_k^{(p)} e^{-2 \pi i kf} \right|^2 } 
= | \Gamma (f)|^2 S_z (f) 
\ , \eqno (3)$$
where the spectral transfer function, $\Gamma (f)$, is defined by Eq.~(3).
%When the spectrum of the residual errors, $S_z (f)$, is white,
In AR spectral analysis, we assume that the autoregressive filter has removed
all of the time correlation and that $\{ z_i \}$ are independent identically
distributed random variables with variance, $\sigma_z^2$, estimated by Eq.~(2).
The resulting AR spectral estimate is given by Eq.~(3) with
 $S_z(f) = \sgh_z^2$.

Thus,  the AR spectral estimate is a rational
approximation to the actual spectral density function $S(f)$. This type
of low order spectral approximation tends to describe the bulk
characteristics of simple spectra well, but has difficulty resolving
the fine features of the spectrum. 
%These autoregressive spectra models also are called maximum entropy models, 
%all-pole models and maximum likelihood models.
Figure 5 shows AR(10) and AR(20) fits to the spectrum. 
The AR fitted spectra have difficulty  fitting the spectral plateau
between 1450-1600 kHz. By AR(20), 
the reduction in the root mean squared error has virtually
saturated as a function of the filter order.
{\it Using the the autoregressive filter accentuates the transient bursts
because it applies the filter, $\Gamma(f)^{-1}$, to  remove the ambient
spectrum, especially the 1 MHz peak, and therefore accentuates the 
frequency components which are only weakly present in the ambient spectrum.}

Spectral prewhitening$^{13,14}$ was developed by J.~Tukey 
to reduce the spectral variation in the analyzed time series.
Prewhitening uses the AR model as an initial filter and then uses local,
%high resolution 
Fourier-based methods, such as the smoothed periodogram or  multiple taper
analysis to estimate the spectrum of the residual
process, $\{ z \}$. 
Prewhitening implicitly assumes that the autoregressive parameters,
$\alpha_i^{(p)}$, are independent of the measured and residual processes.
In our precursor analysis in Sec.~VI, { we use a prewhitening filter
prior to computing the time-frequency distribution.}

\ \\
%\newpage
\noindent
{\bf IV.~Nonstationary plasma fluctuations during the ELMs}
\ \\

When the plasma fluctuations are stationary stochastic processes, we
can identify the time autocovariance of the fluctuations with the Fourier
transform of the spectral density $S(f)$. In Ref.~9, we compare methods of
estimating the spectral density for stationary plasma fluctuations. 
When the spectral density is changing
slowly in time, the natural generalization
is that of Karhunen processes$^{14-16}$:
$$
x_t = \int_{-1/2}^{1/2} A(f,t)e^{ift} dZ(f)
\ , \eqno(4)$$
where $dZ(f)$ is a white noise process with independent spectral
increments. When $A(f,t)$ evolves slowly with respect to the sampling rate,
we can interpret $x_t$ as an approximately stationary process.
The evolutionary spectrum, $S(f,t) = |A(f,t)|^2$, then
corresponds to the instantaneous value of the spectral density.

To estimate the evolutionary spectrum, first we compute the local Fourier
transform, $y(f,t)$, 
of $x_t$ using a sliding tapered (or multitapered) time window,
and the  one point spectral estimate is $\hat{S}(f,t) = |y(f,t)|^2 $,
which we evaluate on a time-frequency grid.
When the sampling rate is much larger than the characteristic time scale,
we can improve on these point estimates of the evolutionary spectrum by
smoothing the spectral density in time and frequency$^{16}$. 
However, we are not in
this limit due to the burstlike nature of the ELMs and precursors.

A good model for these plasma fluctuations is that the fluctuations consist
of a stationary component and a transient bursting component:
$$
S(f,t) = \bar{S} (f) + A_e (t) S_e (f) + \tilde{S} (f,t)
\ , \eqno(5)$$
where the subscript $e$ denotes ``ELM'', and $\tilde{S} (f,t)$ is
the residual transient spectrum. To estimate $\bar{S} (f)$, $A_e (t)$,
and $S_e (f)$, we begin by computing a multiple taper estimate of
$S(f,t)$. To retain high time-frequency resolution, we use 1000-point
segments with 8 tapers ($w \ \sim 20$ kHz). We subtract the mean
values, $\bar{S} (f)$:  ${S} (f,t) - {\bar{S}} (f)$, and then
estimate $A_e (t) S_e (f)$ by computing the singular value
decomposition of
$$
\hat{S} (f,t) - \hat{\bar{S}} (f) = \sum_k%{k=1}^N 
\lambda_k A_k (t) h_k(f) \ . \eqno(6)$$
The singular value decomposition divides the spectrum into its fundamental
components$^{24}$. Neither $A_k (t)$ nor $h_k (f)$ need be positive.
We equate $A_e (t) h_e (f)$ with $\lambda_1 A_1 (t)S_1 (f)$.

In practice, we find that the quasi-coherent part of the 1 MHz peak
has a different time evolution than the rest of the spectrum. To remove
the effect of the IF~frequency, we band limit the signal to exclude
the 1 MHz peak $\pm$ 40 kHz prior to computing the singular value
decomposition. Figure 6 displays the mean
spectral estimate, where we have used 1000 data points corresponding to
a 0.4 msec~time interval with 8 tapers.

The mean spectrum is similar to, but broader than, the multitaper estimate
in Fig.~2. The broadening occurs because we have lowered the frequency
resolution to increase the time resolution.  
Figure 7 presents the first singular time vector, $A_1(t)$, which 
corresponds to the ELM bursts. The actual rise time of the
ELMs is much sharper than that displayed in Figure 7. As we increase the
time resolution, the rise time of the ELMs decreases with the time window
length. Therefore the rise time is not resolved. 

Figure 8 displays the corresponding first  singular
frequency vector, which is broader and more symmetric than the
mean spectrum. The broader and
more symmetric spectrum was originally observed in Ref.~12. However, the
time-frequency singular value decomposition allows us to quantify the
spectrum of a bursting event.
$S_e (f)$, as estimated by the singular value decomposition, is essentially a
time-weighted average of the spectra of the 14 ELMs with the weighting
function given by $A_e (t)$.

Our singular value
decomposition differs from that of Nardone$^{25}$ and Zohm et al.$^{26}$ 
because our two axes are
time and frequency. In contrast, Refs.~25 \& 26 have multiple measurement
channels, and compute a singular value decomposition with time and 
space/mode number as the two axes.

\ \\

%\newpage
\noindent
{\bf V. Spectral evolution  between ELMs}
\ \\

We now  examine the temporal evolution of the spectral density between 
ELMs. We estimate the average growth rate as a function of frequency.
Our results are strictly for discharge \# 49035. Other discharges
sometimes have significantly different spectral evolution, both
quantitative and qualitative.

In our initial analysis, we computed the evolutionary spectrum
as a sequence of spectral estimates. Due to the broad-banded nature
of the plasma turbulence, we present only the total spectral energy
in four 300 kHz bands versus time. By integrating in frequency, we
reduce the variance of the estimate and display only the essential features
of the data. We then estimate the growth rates of the spectrum
during the time after the end of an ELM.

To  estimate 
$\bar{S} (\bar{f},t) = 
\int_{\bar{f}-150 kHz}^{\bar{f}+150 kHz}S(f,t) df $, 
we estimate $S(f)$ on a 0.4 msec interval
centered at $t$ using a 100 taper estimate with $w$ = 150 kHz. 
%We then integrate this estimate in frequency. 
To further reduce the variance, we average
$log_{10}[\hat{\bar{S}} (\bar{f},t)]$ over all 14 quiescent periods.
% between the 14 ELMs. 
This averaging is necessary because the fluctuation level is highly variable,
indicating that there are only one or two waves present at each frequency.
We use the logarithm to reduce the influence of the outliers$^{9,19}$. 
This ELM averaging is reminiscent of sawtooth averaging in heat pulse 
propagation$^{27}$. In sawtooth averaging, the signal is averaged directly to 
reduce the noise while in ELM averaging, we average the log-spectrum to 
determine the spectrum of the noise.

Both the ELM duration and the length of time between ELMs varies from
ELM to ELM. (See Table 1.) We find that the ELM duration is 1.2  $\pm 0.3 $  
msec~and that the quiescent time between ELMs is $6.1\pm 1.5$ msec.
After the 10th ELM, a large amplitude fluctuation burst 
occurs at $t=79.15$ centered at 450 kHz. The length of time between 
the 10th and 11th ELM is
considerably larger, 9.45 msec~versus 5.8 msec. We believe that
{\it this ion fluctuation burst indicates the release of 
some of the free energy
which causes the ELM instability, and thereby delays the onset of the 11th
ELM.} Excluding the time between the 10th and 11th ELM gives the
typical time length between the end of the $(k-1)$th ELM and the beginning of
the $k$th ELM to be 5.8 $\pm $ 1.1 msec.

Since the 14 quiescent periods vary in length,
we need to standardize the intervals to all be of the same length 
%quiescent periods
prior to averaging the spectral estimates. 
%ELM periods because the time between ELMs varies.
%We standardize the quiescent periods to all have the same length.
With this standardization, the spectral evolution is clearly exponential.
Figure 9 displays the mean value of  $log_{10}[\hat{\bar{S}}] (\bar{f},t)$ 
after all quiescent periods are standardized to a length of 5.8 msec.
The  fluctuations in the electron drift direction increase rapidly in the 
first 0.5 msec~after the ELM subsides. During the first half millisecond,
the spectrum between 1500 \& 1800 kHz grows by 50 percent and
the spectrum between 1200 \& 1500 kHz grows by 80 percent.
During this time, the ``ion'' fluctuations actually decrease slightly.

To illustrate this spectral evolution, Fig.~10 plots the spectrum
for three time slices, 0.1, 0.3, and 0.5 msec after the second ELM.
The spectra were calculated with 8 tapers on 1000 data point segments,
and have a frequency resolution of 20 kHz and a time resolution of
0.1 msec. The 1 MHz peak decreases in amplitude and in width. This narrowing
of the 1 MHz line results in a reduction of the ``ion'' spectrum.
The ``electron'' spectrum between 1600 and 1800 kHz is growing noticeably.
In detector \#2, the secondary maximum does not grow appreciably
after the end of the ELM.

Figure 11 plots the corresponding spectral evolution for detector \#2.
The ``ion'' spectrum hardly grows during the initial 0.8 msec
after the ELM. The growth rates of the ``electron'' spectrum for detector \#2  
are roughly half
of those of detector \#3. But the initial phase of rapid  growth lasts
for nearly a full millisecond in detector \#2 while the the rapid growth
period in detector \#3 is only half a millisecond long. 
Thus, the  total  growth of the ``electron'' spectrum in the two detectors is 
nearly the same. In the first millisecond, the ``electron'' spectrum grows  
by 50 \% for $f'$ in 1500 -- 1800 kHz and 
by 100 \% for $f'$ in 1200 -- 1500 kHz.
(We continue to use $f' \equiv$ 1MHz -- $f$ for detector \#2.)
This suggests that the saturated amplitude is a weak function of the initial
growth rate.

Estimating these initial growth rates, 
$\gamma$, with $S(t) \sim \exp(2\gamma t)$, we find that 
for detector \#3, $\gamma = 0.5$ msec$^{-1}$ for  $f$ in 1200 -- 1500 kHz,
and $\gamma = 0.37$ msec$^{-1}$ for  $f$ in 1500 -- 1800 kHz;
for detector \#2, $\gamma = 0.28$ msec$^{-1}$ for  $f'$ in 1200 -- 1500 kHz,
and $\gamma = 0.24$ msec$^{-1}$ for  $f'$ in 1500 -- 1800 kHz.
After this  initial phase, the ``electron'' growth rates slow  by a  factor
of three to  five in detector \#2 and by even less in detector \#3.

During the 5.3 millisecond time period beginning 0.5 milliseconds after
the end of the ELM, 
the level of the electron drift fluctuations
grows slowly. In contrast, {\it the fluctuation level in the
ion drift direction increases by  a factor of five in the
500-800 kHz range and by  a factor of eight in the
200-500 kHz range.}  
At the time of the onset of the ELM, the electron
drift spectrum in 1200 \& 1500 kHz range is roughly 50 \% larger than   
the corresponding ion drift fluctuations.
For detector \# 2, the increase in the ``ion'' spectrum is less:
a factor of 1.6 $\times$ in the 500-800 kHz range 
and by nearly a factor of 2.2 $\times$ in the 200-500 kHz range.
The ratio of the total spectral energy in the ion drift direction  to 
that of the energy in the electron drift direction is much less
for detector \#2 than for detector \#3. This occurs because the electron
is larger in detector  \#2 and the ion energy growth between ELM is less.

Beginning 0.5--1.0 millisecond after the  ELM, 
the evolution of the ``ion'' spectrum
%is nearly linear on the logarithmic scale (see Figs.~9 \& 11),
enters into an exponential growth phase. During this later phase, we find
$\gamma = 0.2$ msec$^{-1}$ for  $f$ in 200 -- 500 kHz,
and $\gamma = 0.15$ msec$^{-1}$ for  $f$ in 500 -- 800 kHz
for detector \#3.
For detector \#2, $\gamma = 0.1$ msec$^{-1}$ for  $f'$ in 200 -- 500 kHz;
for  $f'$ in 500 -- 800 kHz, $\gamma = 0.17$ msec$^{-1}$ for the time
0.3 to 1.3 msec after the ELM and  
$\gamma = 0.01$ msec$^{-1}$ for later times. In summary,
the ``ion'' growth rates are smaller by a  factor of 2.5 -- 3.0 times
the initial ``electron'' growth rates.
We stress that these growth rates vary appreciably from one discharge
to another.

{\it When the ``ion'' spectrum grows more slowly, 
the onset of the ELM is delayed}.
This suggests that the ELM is triggered when the ``ion'' spectrum near the edge
reaches a  critical level. By averaging $log_{10}[{\bar{S}} (\bar{f},t)]$
over 0.44 msec, we find that the critical levels
are $log_{10}[{\bar{S}} (350{\rm kHz},t)] = 2.40  \pm 0.14$ 
and $log_{10}[{\bar{S}} (650{\rm kHz},t)] = 3.45 \pm 0.14$. (See Table 2.)
This corresponds to relative variations of 40\%  in the critical level.
This variation seems large, but ${\bar{S}} (\bar{f},t)$ increases by a factor
of  five to ten. Thus, the variance in the critical level is small relative 
to the total growth in the ``ion'' spectrum.
%However, this ``critical level'' varies by a factor of two.  
The onset of the ELM
need not be caused by the increased fluctuation level. Instead, 
both phenomena may be caused  by the same  destabilizing mechanism.

Standardizing the 14 interval lengths is successful because the final level
of the ion fluctuations varies less than either the length of the 
quiescent interval or the exponential growth rate of the ``ion'' spectrum. 
If we believe that the interesting spectral evolution occurs immediately
after the end of the ELMs, we should align the $k$th quiescent period
at the end of $(k-1)$th ELM. If we believe that the interesting evolution 
occurs immediately prior to the start of the ELMs, 
we should align the $k$th quiescent period at the beginning of the $k$th ELM. 
We tried both of these normalizations; however, no clear pattern emerged
and the ELM to ELM variance was larger than when interval
lengths were standardized.
%of the 13 intervals are not  standardized to be the same length, much more
%complicated dependencies result.

We estimate the variance
of $log_{10}[\hat{\bar{S}}]$ using the empirical variance estimate from the 14 
independent subsequences.
We find that the empirical standard deviation, $\sigma(\bar{f},t)$, of 
$log_{10}[\hat{\bar{S}} (\bar{f},t)]$ varies little in time.
For $\bar{f} = 1350$ and 1650 kHz, $\sigma(\bar{f},t) \sim 0.1$, which 
corresponds to a relative variation of 25 \%. 
For $\bar{f} =$ 650 kHz, $\sigma(\bar{f},t) \sim 0.15$, which 
corresponds to a relative variation of 40 \% and
for $\bar{f} = 350$ kHz, $\sigma(\bar{f},t) \sim 0.19$, which 
corresponds to a relative variation of 55 \%.
These values of $\sigma(\bar{f},t)$ are for detector \#3.
This variation is small relative to the total increase in the 
ion  fluctuations over the 6 milliseconds.
Furthermore, the variance in the mean is $1/14$ of $\sigma^2(\bar{f},t)$.

For detector \#2, $\sigma(\bar{f}',t) \sim 0.1$ for  
$\bar{f}' = 350$ kHz and  650 kHz. Thus the ``ion'' spectrum of detector \#3 
is more variable than  that of detector \#2.
This enhanced variability is partially due to precursor 
activity, which we analyze in Sec. VI.
The empirical variance is much larger than the theoretical  estimate
based on Gaussian statistics because it includes  the ELM to ELM variation
due to random effects including precursor times. 
These effects are random in the sense that the vary randomly from ELM to ELM.
In previous works, we have included the 
effect of tokamak to tokamak variation on energy confinement$^{28}$ and
the effect of discharge to discharge variation on the temperature profile 
shape$^{29}$. 

\ \\

%\newpage
\noindent
{\bf VI. Time-freqency identification of ELM Precursors}
\ \\

In examining the autoregressive residual fit errors, we identified a number
of very short nonstationary bursts in the one and a half millisecond interval
prior to each ELM. The typical duration of the precursor burst 
is roughly 10 $\mu$sec~long. The frequency resolution
is only 75 kHz due to short length (N = 100) of the segment. 
Due to its short life, the concept of a spectrum for the precursor is somewhat
tenuous.

Figure 12 plots the spectrum of one of the longest ($11 \mu$sec) and
most prominent precursors from detector \# 3. 
The precursor is centered at 500 kHz, 
and occurs 0.45 msec~before the second ELM.
The dashed line in Fig.~12  is %Figure 9 provides 
the corresponding estimate on a data segment taken 40 
$\mu$sec~after the
precursor corresponding to the ambient spectrum. 
In Figure 12, we have used three tapers with a frequency half
bandwidth of 75 kHz and then kernel smoothed over an additional 50 kHz
half-width. 

In comparing the two spectra, we see that the spectrum in the 400-600 kHz
range is not only much larger than its normal size, but also larger than the
fluctuations in the 1450-1750 kHz range, and this is not typical. 
%If the 1 MHz IF~frequency were the zero frequency, then 
The ambient fluctuations in the 1450-1750 kHz range correspond to 
electron drift direction while
the precursor frequency of 500 kHz  corresponds to ion drift.
%waves. However the plasma is rotating at an unknown rate and this shifts the
%frequency in the plasma rest frame by some unknown amount. Thus, all we can
%say for certain is that the precursor frequency is shifted by roughly 1 MHz
%relative to the ambient spectrum.

To better quantify the occurrence rate, the duration and strength of the
precursor, we have computed the time-frequency distribution 
for the 4 milliseconds prior to each ELM. 
To have a very short time resolution, we use only two tapers 
and smooth over 100 kHz.
We compute the mean, $\bar{S} (f)$, and standard
deviation, $\sigma (f)$, of the time-frequency distribution averaged over
time for each frequency. We standardize the time-frequency plot by
subtracting off the mean and dividing by the standard deviation,
so that the resulting function is
$${S_z (f,t) - \bar{S}_z (f) \over \sigma (f)}\ .$$
%Figure 10 displays a gray scale plot of 

We identified two types of short bursts: bursts
which are centered in the 250-650 kHz range, and bursts which have a
substantial part of their energy content in the 1 MHz range. After inspecting
the fourteen time-frequency plots, we concluded that the bursts centered at
1 MHz did
not have any noticeable correlation with the onset of the ELMs while the low
frequency bursts occur almost exclusively in the one and a half milliseconds
prior to the ELM. Furthermore, the
bursts in the 1 MHz range occur at a lower amplitude
($10 \sigma(f)$)   than those in the 250-650 kHz range ($30 \sigma(f)$).

To enhance the low frequency fluctuations and to filter out the ambient
spectrum centered at 1 MHz, we prewhiten the time series by applying a tenth
order autoregressive filter prior to computing the time-frequency
distribution. Prewhitening reduces the amplitude of the bursts 
in the 1 MHz range. 
For graphical effect, we have set all values of the standardized 
time-frequency distribution which are less than three to exactly zero.

Figure 13 displays a three-dimensional plot of the prewhitened transformed  
time-frequency distribution for the four milliseconds~immediately prior
to the second ELM. The precursor burst at $t =$ 12.734 msec~
is clearly visible and is centered at 500 kHz.
Figure 14 presents a similar time-frequency plot
%distribution of the prewhitened data
for the 4 millisec~time interval just prior to first ELM. 
The large amplitude, 10 $\mu$sec~precursors burst are visible at 
$t =$  5.186 and  5.56 msec.
Only the first burst is visible on the AR residual plot of Fig.~3b.
The precursors bursts in the first two ELMs are localized around 500
kHz. However, the frequency range of the precursors for later ELMs varies
between 250 kHz and 650 kHz.   

Figure 15a plots the integrated energy in the frequency band, $[300 -700]$ kHz,
versus time. The precursor bursts at $t =$  5.186 and  5.56 msec
and at $t =$ 12.734 msec are clearly visible and above the noise.
The ELM oscillations begin at $t =$  6.07 and  13.19 msec.  
Figure 15b gives the corresponding time history in the
$[900 -1100]$ kHz band. Many fluctuations occur in this frequency band,
and there appears to be no correspondence between these fluctuations
and the imminent onset of ELM activity. 
Figure 15c gives the  time history in the $[1300 -1700]$ kHz band,
corresponding to electron drift direction. No evidence of precursor 
activity is indicated. The spectral intensity in this frequency range
is virtually constant between ELMs. Thus,
Fig.~15 shows  that the precursor bursts are localized
in the $300 -700$ kHz band. %  which corresponds to ion drift directions.

We set a threshold of $28 \sigma(f)$ on the standardized prewhitened 
time-frequency estimates. We then record all of the events which exceed
this magnitude  during the four milliseconds immediately prior to each ELM.
The threshold value of $28 \sigma(f)$ was chosen to give the best 
identification rate for the ELMs. 
We find that the amplitudes of the instantaneous bursts
have an approximately two-humped distribution, i.e. very few bursts occur
with amplitudes of 18--30 $\sigma(f)$. Thus, we can change the threshold
from 15$\sigma(f)$ to 30$\sigma(f)$ and only slightly reduce the success
rate of predicting ELMs.

Table 3 summarizes our findings on the ELM precursor bursts. 
Column 2 gives the time until the next ELM. 
Column 3 gives the maximum burst amplitude in $\sigma(f)$ and
Column 4 gives the characteristic frequency of the burst. 
Column 5  gives the duration in $\mu$sec.
The seventh ELM has a number of bursts in the 250-650 kHz, but they
have relatively small amplitude and therefore do not exceed the
threshold. The  eleventh ELM has a high amplitude burst at 1350 kHz instead of
in the  300-600 kHz range. The
large amplitude, low frequency burst  after the 10th ELM 
is not listed in Table 3 because it occurs more
than 4 milliseconds~before the next  ELM.

The shorter duration (2-11 $\mu$sec) of  these precursors 
is probably due to a combination of physics differences, 
the fast sampling rate on the TFTR diagnostic, and our high resolution
estimation procedure. {\it The short duration of the precursor events 
reduces their physical significance as a cause of the ELMs and probably
means that they are more of a symptom of a change in plasma conditions
as the stability boundary is approached.} We also caution that our analysis 
only applies to this particular TFTR discharge and that other discharges may 
be different.

These short precursor bursts occur only in detector \#3, and we have
found no corresponding phenomena in detector \#2. One explanation
is that the precursors are occuring at the plasma edge and that the
scattering volume for detector \#2 is  too far in the plasma interior
to sense the precursors. Alternatively,
the absence of precursors in detector \#2 could be due to the difference 
in measured $k$ values of the two detectors.

We have examined several other discharges, and the phenomena of high frequency 
bursts in the ion drift direction appears to be  robust. In several of the
other discharges, the precursors were weaker and less reliable.

\ \\

%\newpage
\noindent
{\bf VII. Summary}
\ \\

In this article, we described several techniques which we have found
useful in examining the nonstationary bursts: autoregressive filters to isolate
nonstationary phenomena, prewhitening to reduce bias, and the singular value
decomposition to isolate the fundamental components of the signal,
including the spectrum during the ELMs. We have used a high resolution smoothed
multitaper estimate of the evolutionary spectrum to isolate short-lived
high frequency ELM precursor bursts. Our evolutionary spectra estimate uses
many fewer degrees of freedom than in our previous work for two reasons:
first, the bursts are so transient that segment lengths of more than
100 points blur the time resolution; second, the events appear to be
coherent (i.e. consist only of a single mode), and thus
are easier to estimate than an ensemble of transient modes.

Our main experimental findings are: 1) the ELM spectrum is more symmetric
and broader than the stationary spectrum; 2) the existence of high frequency
ELM ``ion'' precursor bursts; 3) the ``electron'' spectrum returns to its
ambient value within a half a millisecond~after the ELM while the ``ion'' spectrum
grow by a factor of eight between ELMs; 4) the onset of the ELMs 
appears correlated with the level of ion fluctuations (for detector \#3);
5) growth rates of 0.1--0.5 msec$^{-1}$ are estimated after the end of an ELM.
Results 3-5 are not always present in other discharges. In some discharges,
the ``electron'' spectrum grows between ELMs and the ``ion'' spectrum is
saturated. 

Figure 9 shows that the ``ion'' spectrum grows exponentially
between ELMs. When the ion fluctuation spectrum increases more slowly,
the onset of the ELM is delayed. It appears that when the ``ion'' spectrum
reaches a  critical level, the ELM is triggered. However, this
``critical level'' varies by 40 \%.  The onset of the ELM
need not be caused by the increased fluctuation level. Instead, 
both phenomena may be caused  by the same  destabilizing mechanism.
Some other discharges have a fixed level for the ion spectrum, and 
have no indication of  an ion character to the ELM onset.

The electron drift spectrum also appears to grow exponentially to a critical
level and then enters a slow growth phase. 
The initial ``electron'' growth  rates are 
faster in detector \#3 than in detector \#2, but the length of time of 
rapid growth is longer in detector \#2; thus the total increase in the 
electron spectrum is the nearly the same in both detectors.
{\it Because the spectrum grows exponentially  between ELMs, we interpret this
as the linear phase of the plasma instabilities and the 0.1-0.5 msec$^{-1}$
growth rates to be the linear instability growth rates}.  
Our interpretation is based on the assumption that nonlinear saturation
is an algebraic function of time. Similarly, if the turbulence level were
evolving slowly due to changing plasma conditions, such as changes in
the electric field$^{20-21}$, the spectrum would almost certainly be evolving 
more slowly than exponentially.

%The precursor bursts have been reported previously$^{2,4,12}$,
%but we believe that the techniques which we have just described have
%resulted in more quantitative and higher resolution estimates. 
The precursor
bursts, which we identify, have much shorter lifetime than previously
reported measurements$^{2,4}$, and are intermittent.
Their high frequency, ion drift nature further suggests than the ELM
onset is related to increased ion turbulence.   
%could be associated with ion drift frequencies. 
Intermittent  fluctuation bursts occur at other frequencies,
especially near 1 MHz. However, these bursts occur at a lower level 
($10 \sigma(f)$ instead of $30 \sigma(f)$), and are not correlated with the 
onset of an ELM burst.  
%The main experimental qualifications to our results are a)
%we have examined only one discharge with a single diagnostic, b) the plasma
%is rotating at an unascertained rate, thereby making the precursor
%frequency in the plasma reference frame somewhat uncertain. In closing, 
We note that the precursors on this TFTR discharge occur
with sufficient regularity that the occurrence of the next ELM can be 
predicted with about 80\% accuracy. For this to be of use in fusion reactor 
design, we must hope that the  same precursor signature occurs in 
other devices and in most discharges.

\ \\
\noindent
{\bf Acknowledgements}

We thank  R.~Nazikian for many interesting discussions which 
stimulated this work and for providing the data used in this analysis. 
%detailed explanations of the physics of the 
%T.F.T.R. microwave scattering experiment.
%KSR thanks C.~Hurvich for 
%extensive discussions and advice on time series analysis,
%and W.~Sadowski and H.~Weitzner for their support.
% has benefited from extensive discussions with  C.~Hurvich
%KSR has benefited from many discussions with C.~Hurvich. 
%and Craig Lindberg. 
We thank the TFTR group for  allowing us to analyze
the microwave scattering data. 
%and we thank Alan Chave for providing his multitaper time series code. 
%Norton Bretz and Raffi Nazikian provided the TFTR data.
The work of KSR and AS was funded by the U.S. Department of Energy,
Grant No.~DE-FG02-86ER53223.

%I thank C. Hurvich for discussions on data based selection criteria.
%The valuable comments of the referee are also appreciated.
%referee suggested several valuable extension to the original
%manuscript, wh
%This work was supported by the U.S. Department of Energy. 
%Grant No.DE-FG02-86ER-53223.

\ \\

\np
{\bf Bibliography}{}
\begin{enumerate}
%\begin{bibliography}{}

\item ASDEX team,
Nuclear Fusion {\bf 39},  1959 (1989).

\item 
C.~Bush, N.~Bretz, R.~Nazikian, B.C.~Stratton, E.~Synakowski,
G.~Taylor, R.~Budny, A.T.~Ramsey, S.D.~Scott, M.~Bell, R.~Bell, H.~Biglari,
M.~Bitter, D.S.~Darrow, P.~Efthimion, R.~Fonck, E.D.~Fredrickson,
K.~Hill, H.~Hsuan, S.~Kilpatrick, K.M.~McGuire, D.~Manos, D.~Mansfield,
S.S.~Medley, D.~Mueller, Y.~Nagayama, H.~Park, S.~Paul, S.~Sabbagh,
J.~Schivell, M.~Thompson, H.H.~Towner, R.M.~Wieland, M.C.~Zarnstorff,
and S.~Zweben, ``Characteristics of the TFTR limiter H-mode: the transition,
ELMs, transport and confinemeent,''
Submitted for publication in Nuclear Fusion.
%.~{\bf 61}, 3031 (1990).

\item S.M.~Kaye, J.~Manickam, N.~Askura, R.E.~Bell, Yun-Tung Lau,
B.~LeBlanc, C.E.~Kessel, H.W.~Kugel, S.F.~Paul, S.~Sesnic, H.~Takahashi,
Nuclear Fusion {\bf 30},  2621 (1990).

\item K.~McGuire, V.~Arunasalam, C.W.~Barnes, M.G.~Bell, M.~Bitter, R.~Boivin,
N.L.~Bretz, R.~Budny, C.E.~Bush, A.~Cavallo, T.K.~Chu, S.A.~Cohen,
P.~Colestock, S.L.~Davis, D.L.~Dimock, H.F.~Dylla, P.C.~Efthimion,
A.B.~Ehrhardt, R.J.~Fonck, E.~Fredrickson, H.P.~Furth, G.~Gammel,
R.J.~Goldston, G.~Greene, B.~Grek, L.R.~Grisham, G.~Hammett,
R.J.~Hawryluk, H.W.~Hendel, K.W.~Hill, E.~Hinnov, D.J.~Hoffman, J.~Hosea,
R.B.~Howell, H.~Hsuan, R.A.~Hulse, A.C.~Janos, D.~Jassby, F.~Jobes,
D.W.~Johnson, L.C.~Johnson, R.~Kaita, C.~Kieras-Phillips,
S.J.~Kilpatrick, P.H.~LaMarche, B.~LeBlanc, D.M.~Manos, D.K.~Mansfield,
E.~Mazzucato, M.P.~McCarthy, M.G.~McCune, D.H.~McNeill, D.M.~Meade,
S.S.~Medley, D.R.~Mikkelsen, D.~Monticello, R.~Motley, D.~Mueller,
J.A.~Murphy, Y.~Nagayama, R.~Nazikian, E.B.~Neischmidt, D.K.~Owens,
H.~Park, W.~Park, S.~Pitcher, A.T.~Ramsey, M.H.~Redi, A.L.~Roquemore,
P.H.~Rutherford, G.~Schilling, J.~Schivell, G.L.~Schmidt, S.D.~Scott,
J.C.~Sinnis, J.~Stevens, B.C.~Stratton, W.~Stodiek, E.J.~Synakowski,
W.M.~Tang, G.~Taylor, J.R.~Timberlake, H.H.~Towner, M.~Ulrickson,
S.~von Goeler, R.~Wieland, M.~Williams, J.R.~Wilson, K.-L.~Wong,
M.~Yamada, S.~Yoshikawa, K.M.~Young, M.C.~Zarnstorff, and
S.J.~Zweben,
Phys.~Fluids B {\bf 2},  1287 (1990).

\item E.J.~Doyle, R.J.~Groebner, K.H.~Burrell,  P.~Gohil, T.~Lehecka,
N.C.~Luhmann, Jr., H.~Matsumoto, T.H.~Osborne, W.A.~Peebles, and R.~Philipona, 
Phys.~Fluids B {\bf 3}, 2300 (1991).

\item H.~Zohm, F.~Wagner, M.~Endler, J.~Gernhardt, E.~Holzhauer,
W.~Kerner, V.~Mertens, 
Nuclear Fusion {\bf 32},  489 (1992).

\item C.~Bush, J.~Schivell, G.~Taylor, N.~Bretz, A.~Cavallo, E.~Fredrickson,  
A.~Janos, D.~Mansfield, K.~McGuire, R.~Nazikian, H.~Park, A.T.~Ramsey,
B.C.~Stratton, and E.~Synakowski,
Rev.~Sci.~Instr.~{\bf 61}, 35 (1990).
%.~{\bf 61}, 3031 (1990).

\item D.J.Grove and D.M.~Meade,
Nuclear Fusion {\bf 25},  1167 (1985).

\item K.S.~Riedel,  A.~Sidorenko, D.J.~Thomson,
``Spectral estimation of plasma fluctuations I: Comparison of methods,''
Published in this issue, Physics of Plasmas {\bf 1} page ? (1994).

\item N.~Bretz, P.~Efthimion, J.~Doane, and A.~Kritz, Rev.~Sci.~Instr.
{\bf 59},  1538 (1988).

\item N.~Bretz, R.~Nazikian, W.~Bergin,  M.~McCarthy,
Rev.~Sci.~Instr.~{\bf 61}, 3031 (1990).

\item N.~Bretz, R.~Nazikian, and  K.~Wong, {\it Proceedings  of the
17th European Phys.~Soc.~Conf.}, 
(European Phys.~ Soc., Amsterdam, 1990) p.~1544.

%\item R.~Nazikian, N.~Bretz,  E.~Fredrickson, Y.~Nagayama, E.~Mazzucato, 
%K. McGuire, H.K.~Park, G.~Taylor, A.~Cavallo, M.~Diesso, J.~Felt,
%{\it Proceedings  of the 18th European Phys.~Soc.~Conf.}, 
%(European Phys.~Soc., Berlin  1991) Vol. I p. 265.

\item{P.J.~Brockwell and R.A.~Davis,  %(1981).~
{\it Time series - theory and methods.} 
({Springer Verlag}, New York 1991).~ }

\item{M.B.~Priestley,  %(1981).~
{\it Spectral analysis and time series.} 
({Academic Press}, New York 1981).~ }

\item 
{ M.B.~Priestley, 
%{ Evolutionary spectra and nonstationary processes.} 
{ J.~Roy.~Stat.~Soc.~Ser.~B} {\bf 27}, 204 (1965).~}

\item 
{K.S.~Riedel, 
%{ Optimal kernel estimation of evolutionary spectra.} 
{I.E.E.E.~Trans.~on Signal Processing} {\bf 41}, 2439 (1993).~}

%\item{D.J.~Thomson (1977).
%{Spectrum estimation techniques for characterization
%and development of the WT4 waveguide}
%{\it Bell Syst.~Tech.~J.}, {\bf 56}, Part I: 1769-1815.}  

\item
{D.J.~Thomson, 
%{Spectrum estimation and harmonic analysis}
{\   Proc.~I.E.E.E.} {\bf 70}, 1055 (1982).}  

\item
{J.~Park, C.R.~Lindberg, and F.L.~Vernon,   %(1987).
%{Multitaper Spectral analysis of high frequency seismiograms} 
{\   J.~Geophys.~Res.} {\bf 92B}, 12,675 (1987).}

\item
{D.J. Thomson and A.D. Chave,  %(1990).
%{Jackknife error estimates for spectra, coherences and transfer functions} 
in {\it Advances in spectrum analysis},
edited by S. Haykin, (Prentice-Hall, New York 1990) 
Ch. 2,  pg. 58-113.}

\item  R.J.~Groebner, K.H.~Burrell,  and R.P.~Seraydarian,
Phys.~Rev.~Lett.~{\bf 64}, 3015 (1990).

\item  R.J.~Groebner, W.A.~Peebles, K.H.~Burrell, T.N.~Carlstrom, P.~Gohil, 
R.P.~Seraydarian, E.J.~Doyle, R.~Philipona, 
H.~Matsumoto, B.~Cluggish,
{\it Proceedings  of the Thirteenth Conference on  
Plasma Physics and Controlled Fusion Research}, IAEA-CN-531A-VI-4
(International Atomic Energy Agency, Vienna  1991) p.~453.

%\item  R.J.~Groebner, K.H.~Burrell, ?? P.~Gohil, T.~Lehecka,
%N.C.~Luhmann, Jr., H.~Matsumoto, T.H.~Osborne, W.A.~Peebles, 
%and R.~Philipona, Rev.~Sci.~Instr.~{\bf 61}, 2920 (1990).

\item
{B.~Kleiner, R.D.~Martin and D.J.~Thomson, 
%{Spectrum estimation and harmonic analysis}
{\  J.~Royal Stat.~Soc.~Series B}, {\bf 41}, 313 %351 
(1979).}  

\item
{R.D.~Martin and D.J.~Thomson, 
%{Spectrum estimation and harmonic analysis}
{\ Proc.~I.E.E.E.}, {\bf 70}, 1096 (1982).}  

\item
{D.J. Thomson, %
%{Quadratic inverse spectrum estimates: applications to paleoclimatology}
{\   Phil. Trans. R. Soc. Lond. A}, {\bf 330}, 601%616 
(1990).} %-597. }  

\item C.~Nardone,  Plasma Physics and Controlled
Fusion, {\bf 34},  1447 (1992)

\item H.~Zohm, J.M.~Greene, L.L.~Lao, \& E.J.Strait,
``Mirnov coil analysis in the D-IIID tokamak using the
singular value decomposition method.'' G.A.~Report GA-A20886,
Submitted to Nuclear Fusion, (1992).
%Plasma Physics \& Nuclear Fusion ?{\bf 25},  1167 (1985).

\item  K.S.~Riedel, E.~Eberhagen, O.~Gruber, K.~Lackner, G.~Becker,
O.~Gehre, V.~Mertens, J.~Neuhauser, F.~Wagner, and the  ASDEX team,
{Nuclear Fusion} {\bf 28}, p.~1509
(1988).

\item  K.S.~Riedel, 
%``Tokamak to tokamak variation and collinearity in scaling laws,''
Nuclear Fusion {\bf 30},  755, (1990).

\item 
P.J.~McCarthy,  K.S.~Riedel,  O.J.W.F.~Kardaun,  H.~Murmann, K.~Lackner,
Nuclear Fusion {\bf 31},  1595, (1991)

\end{enumerate}
%\end{bibliography}{}

\newpage  
%\begin{center}
\centerline{\em Table 1: ELM and quiescent period duration}
%\end{center}
\begin{tabular}{rccc}
ELM \# & Quiescent time & Duration of ELM & Total period \\
       &     (msec)   &   (msec) & (msec) \\
1 & 4.45 & 1.92 & 6.37 \\
2 & 5.2  & 0.8  & 6    \\
3 & 4.08 & 1.4  & 5.48 \\
4 & 5.38 & 0.8  & 6.18 \\
5 & 7.71 & 1.0    & 8.71 \\
6 & 4.46 & 1.55 & 6.01 \\
7 & 7.21 & 1.1  & 8.31 \\
8 & 5.73 & 1.39 & 7.12 \\
9 & 6.6  & 2.12 & 8.72 \\
10 & 6.3  & 1.26 & 7.56 \\
11 & 9.44 & 1.11 & 10.55 \\
12 & 5.64 & 0.6  & 6.24 \\
13 & 7.19 & 1.14 & 8.33 \\
14 & 5.75 & 0.85 & 6.6\\
Mean & 6.08 (5.82) & 1.22 & 7.3 (7.05) \\
Std.~dev.& 1.46 (1.14) & 0.43 & 1.45 (1.15)\\
\end{tabular}
\vspace{.1in}

The mean and standard deviation in parentheses are computed by excluding 
the interval between the tenth and eleventh ELMs.

\np
\begin{center}
{\em Table 2: Critical ion log-spectrum levels}
\end{center}

\begin{tabular}{rcc}
ELM \# & 200--500 kHz & 500--800 kHz 
\\  1 & 2.19 & 3.54
\\  2 & 2.51 & 3.54
\\  3 & 2.50 & 3.68
\\  4 & 2.59 & 3.41
\\  5 & 2.37 & 3.32
\\  6 & 2.42 & 3.54
\\  7 & 2.30 & 3.30
\\  8 & 2.23 & 3.27
\\  9 & 2.16 & 3.22
\\ 10 & 2.41 & 3.44
\\ 11 & 2.50 & 3.47
\\ 12 & 2.51 & 3.44
\\ 13 & 2.55 & 3.72
\\ 14 & 2.39 & 3.40
\\ Mean & 2.40 & 3.45
\\Std.~dev. & 0.14 & 0.15
\\
\end{tabular}

The critical level of the ion fluctuations is calculated
by averaging $log_{10}[\hat{\bar{S}} (\bar{f},t)]$
over 0.44 msec.
%\vspace{.2in}
\np
\begin{center}
{\em Table 3: Ion precursor bursts}
\end{center}

\begin{tabular}{rcccc}
ELM \# & Time until ELM & Amplitude & Frequency & Duration \\
       &     (millisec)   &   ($\sigma(f)$) & (kHz) & ($\mu$sec) 
\\ 1 & 0.88 & 35 & 300--800 & 2
\\ 1 & 0.51 & 31 & 650 & 2
\\ 2 & 0.45 & 35 & 500 & 11
\\ 3 & 1.14 & 65 & 300 & 6
\\ 3 & 0.80 & 50 & 400 & 4
\\ 3 & 0.56 & 89 & 600 & 7
\\ 4 & 0.37 & 34 & 400 & 2
\\ 5 & 1.04 & 90 & 200--700 & 6
\\ 6 & 0.24 & 36 & 650 & 8
\\ 8 & 1.49 & 42 & 450 & 10
\\ 9 & 1.12 & 52 & 350 & 3
\\ 10 & 1.44 & 19 & 500 & 3
\\ 11 & 0.49 & 29 & 1350 & 2
\\ $12^*$ & .1-.2 & 27 &200- 550 & -
\\ 13 & 1.83 & 37 & 500 & 6
\\ 13 & 0.54 & 37 & 750 & 3
\\ 14 & 0.13 & 43 & 200--600 & 3
\\
\end{tabular}

* No identifiable mode, only broad-banded activity above threshold

%\vspace{.2in}
\includepdf[pages=-,pagecommand={}]{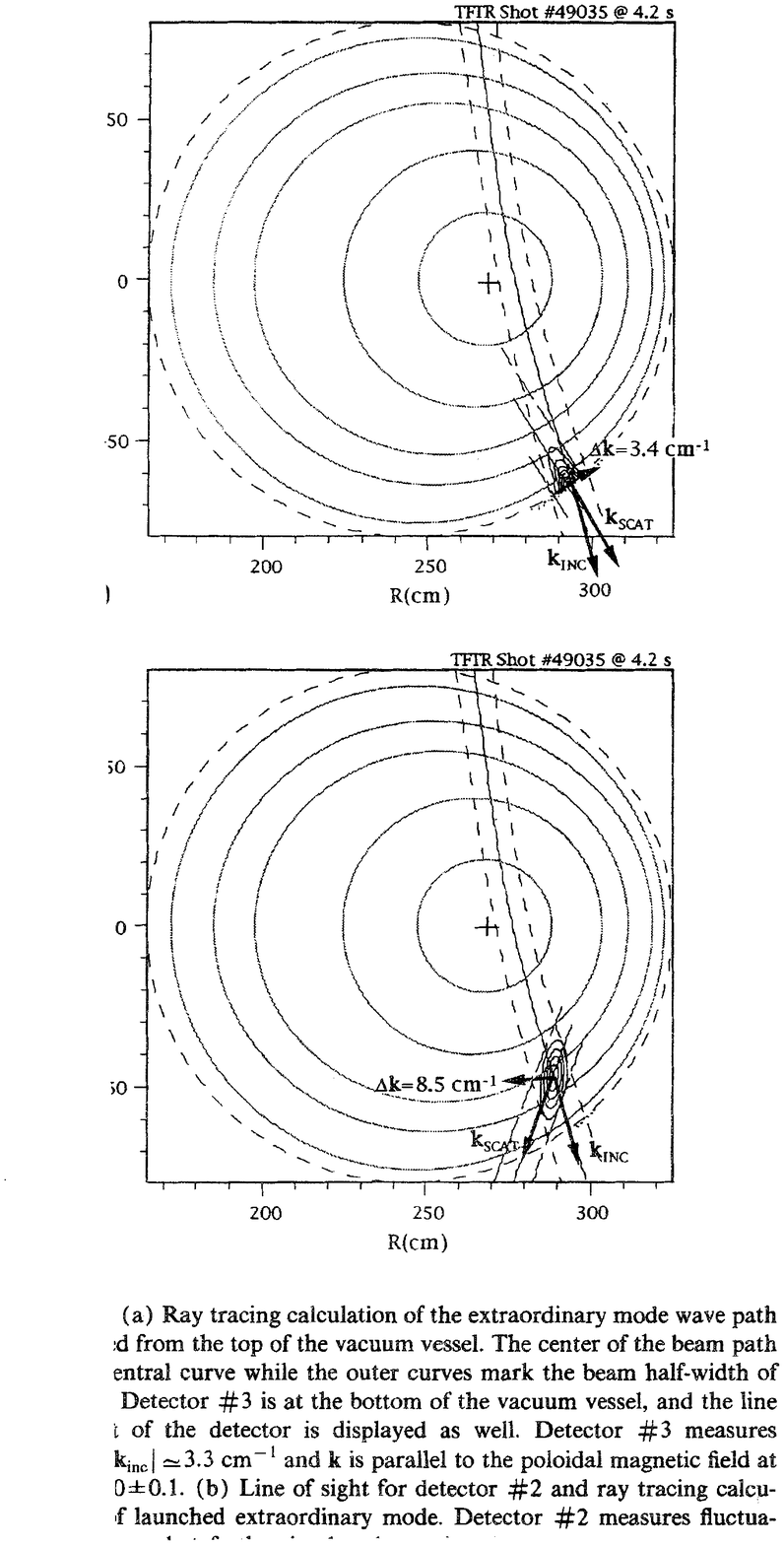}

\newpage
{{  Figure Captions:}
\vspace{.2in}

Figure 1a: Ray tracing calculation of the
extraordinary mode wave path launched from the top of the vacuum vessel.
The center of the  beam path is the central
curve while the  outer curves  mark the beam half-width of 5.0 cm.
Detector \#3 is  at the bottom of the vacuum vessel, and the line
of sight of the detector is displayed as well.
Detector \#3 measures $|\kv_{scat}- \kv_{inc}|\ \siml\  3.3\ {\rm cm}^{-1}$ 
and $\kv$ is parallel to the 
poloidal magnetic field at $\frac{r}{a}\ \siml\ 1.0 \pm 0.1$.
%Figure ?a displays the geometry of the transmitted extraordinary
%mode and the receiver line of sight, the three curves emerging from the bottom
%right are the center of the detector line of sight and its inner and outer
%half-widths.

\vspace{.18in}

Figure 1b: Line of sight for detector  \#2 and  ray tracing calculation
of launched extraordinary mode.
Detector \#2 measures fluctuations somewhat farther in the plasma  interior,
at $\frac{r}{a}\ \siml\ 0.75 \pm 0.1$ with 
$|\kv_{scat}- \kv_{inc}|\ \siml\  8.5\ {\rm cm}^{-1}$.

\vspace{.18in}

Figure 2: Smoothed spectrum of entire 524,288 point segment, estimated with
20 orthogonal tapers  with %sequentual deselection, 
$w = 0.1$ kHz and then kernel smoothed over 20 kHz.~
The central peak at 1 MHz is partially coherent and 
corresponds to the receiver intermediate frequency 
caused by wall and waveguide reflections.
%The broadening of the 1 MHz peak is believed to
%be due to intense edge fluctuations$^{11,12}$ 
%at low $k$, $k_{\perp} \sim 0.1$ cm$^{-1}$. 
%be due to reflection from the vacuum vessel wall. 
The local broadening
from 1450 kHz to 1750 kHz is due to plasma fluctuations which are rotating  
in the electron drift wave direction.
We normalize the spectrum of detector \#2 such that the two spectra have 
the same amplitude when the broadening of the 1 MHz line begins. The ``ion''
spectra are identical while the ``electron'' spectrum of detector \#2
is three times larger.

\vspace{.18in}

Figure 3a: The  12 millisecond data segment from $t = 2$ msec~to 
$t = 14$ msec
of the microwave scattering diagnostic for T.F.T.R.~discharge \# 49035.
The ELMs occur at $t =$  6.07 and 13.19  msec.
A number of fluctuation bursts prior to the ELM are visible; however,
most of these bursts are due to changes in the 1 MHz frequency range.
Two precursor bursts in the $250-650$ kHz range occur at
$t =$  5.186 and  12.734 msec;
however, it is difficult to distinquish these 
bursts from the IF~frequency bursts.

%Raw data detector 1.

\vspace{.18in}

Figure 3b: Filtered data after using a tenth order autoregressive filter.
Since the sawtooth spectrum is broader than the ambient spectrum, the ELM
amplitude is enhanced relative to the background level. 
The intermittent bursts which are concentrated in the 1 MHz range 
are reduced by filtering. In contrast,  
the two precursor bursts in the $300-600$ kHz range, at
$t =$  5.186 and  12.734 msec,~have been enhanced.
%This indicates that the nonstationarity is concentrated in the 1 MHz  peak.

\vspace{.18in}

Figure 4a: Microwave scattering data for TFTR discharge \# 50616.
Sawtooth occurs at $t =  3.86$ msec~and is barely visible. The variance
appears to be growing linearly in time.

%Raw data detector 1.

\vspace{.18in}

Figure 4b: Filtered data after using a tenth order autoregressive filter.
Since the sawtooth spectrum is broader than the ambient spectrum, the effect
of the sawtooth  is greatly enhanced by filtering. The linear growth of the
variance is eliminated by filtering. This indicates that the nonstationarity
is concentrated in the 1 MHz  peak.

\vspace{.18in}
Figure 5: Autoregressive estimate spectrum of entire data set,
estimated by the method of moments.
The 1 MHz peak is  artificially broadened due to model misfit.

\vspace{.18in}

Fig.~6: Mean spectral estimate, $\Sbr(f)$ of the time-frequency distribution,
$S(f,t)$, estimated using 1000 point samples with eight tapers ($w = 20$ kHz)
and then kernel smoothed over  20 kHz. The mean spectrum is broader than 
the estimated spectrum in Fig.~2 because the frequency resolution is lower.

\vspace{.18in}

Fig.~7: First time vector of the singular value decomposition of the
time-frequency distribution. Each of the 14 peaks corresponds to an ELM
burst. The rise time of the ELMs in Fig.~7 is significantly longer than  
the actual rise time because we have reduced the time resolution to increase
the frequency resolution in Fig.~8.
%The time-frequency estimate is explained in the Fig.~6 caption.
We remove the IF~frequency at 1 MHz prior to computing the singular
value decomposition.

\vspace{.18in}

Fig.~8: First frequency vector of the singular value decomposition of the
time-frequency distribution. The spectrum during the  ELM is broader and
more symmetric than the mean spectrum of Fig.~6.

\vspace{.18in}
Fig.~9:
${\bar{S}} (\bar{f},t) \equiv
\int_{\bar{f}-150 kHz}^{\bar{f}+150 kHz}S(f,t) df $
of detector \# 3, averaged over the 14 quiescent periods.
The quiescent periods are standardized to a length of 5.8 msec.
Since the curves  are nearly straight, the growth rates of the 
ion fluctuations are exponential.  When the lengths of the 
13 intervals are not  standardized to be the same length, more
complicated dependencies result.

The  fluctuations in the electron drift direction increase rapidly in the 
first 0.5 msec~after the ELM subsides. %During this half millisec,
%the spectrum between 1500 \& 1800 kHz grows by 40 percent and
%the spectrum between 1200 \& 1500 kHz grows by 75 percent.
%During this time the ``ion'' fluctuations actually decrease slightly.
During the next 5.3 msec, the level of the electron drift fluctuations
is virtually constant. In contrast, {the fluctuation level in the
ion drift direction increases by  a factor of five in the
500-800 kHz range and by a factor of eight in the
200-500 kHz range.}   %At the time of the onset of the ELM, the electron
%drift spectrum in 1200 \& 1500 kHz range is roughly 50 \% larger than   
%the corresponding ion drift fluctuations.
 During this time, the growth rates are
$\gamma = 0.2$ msec$^{-1}$ for  $f$ in 200 -- 500 kHz,
and $\gamma = 0.15$ msec$^{-1}$ for  $f$ in 500 -- 800 kHz.
{ When the ``ion'' spectrum grows more slowly, the onset of the ELM is delayed}.
%Standardizing the 13 interval lengths is successful because the final level
%of the ion fluctuations.

\vspace{.18in}
Fig.~10: Spectrum
for three time slices 0.1, 0.3, and 0.5 msec after the second ELM.
The ``electron'' spectrum between 1600 and 1800 kHz is growing noticeably.
The 1 MHz peak decreases in amplitude and in width. This narrowing
of the 1 MHz line results in a reduction of the ``ion'' spectrum.
In detector \#2, the corresponding secondary maximum does not  grow appreciably
after the end of the ELM.
The spectra were calculated with 8 tapers on 1000 data point segments,
and have a frequency resolution of 20 kHz and a time resolution of
0.1 msec.

\vspace{.18in}
Fig.~11: ${\bar{S}} (\bar{f},t) \equiv
\int_{\bar{f}-150 kHz}^{\bar{f}+150 kHz}S(f,t) df $
of detector \# 2, averaged over the 14 quiescent periods.
In the first millisecond, the ``electron'' spectrum grows  
by 50 \% for $f'$ in 1500 -- 1800 kHz and 
by 100 \% for $f'$ in 1200 -- 1500 kHz.
(We continue to use $f' \equiv$ 1MHz -- $f$ for detector \#2.)
For detector \#2, $\gamma = 0.28$ msec$^{-1}$ for  $f$ in 1200 -- 1500 kHz,
and $\gamma = 0.24$ msec$^{-1}$ for  $f$ in 1500 -- 1800 kHz.
After this  initial phase, the ``electron'' growth rates slow  by a  factor
of  three to five.
The ``ion'' spectrum increases by 1.6 $\times$ in the 500-800 kHz range 
and by 2.2 $\times$ in the 200-500 kHz range.}
In detector \#2, the ``electron'' spectrum
is larger than in detector \#3 and the growth of the ion spectral energy  
between ELM is less:
$\gamma = 0.1$ msec$^{-1}$ for  $f'$ in 200 -- 500 kHz;
for  $f'$ in 500 -- 800 kHz, $\gamma = 0.17$ msec$^{-1}$ for the time
0.3 to 1.3 msec after the ELM and  
$\gamma = 0.01$ msec$^{-1}$ for later times.
Thus the ``ion'' spectrum appears to be more important near the plasma edge.

\vspace{.18in}
Fig.~12: Estimated spectral density during a precursor 0.45 msec~before
the second ELM. The secondary peak at 500 kHz occurs only during 
the precursors.
This precursor is particularly long lived and resolvable.
Spectrum is computed on a 100-point segment using 3 tapers with a 
bandwidth of 75 kHz followed by a kernel smoother with a half-width of 50 kHz.

%\vspace{.18in}

Dashed line: Corresponding estimated spectral density for the 100-point segment
 40 $\mu$sec~later. The precursor peak has totally disappeared and the 
spectrum has returned to its ambient shape. Since the frequency resolution 
of Fig.~12 is less than that of Fig.~5, the spectrum in Fig.~12
is correspondingly broader (just as the spectrum in Fig.~6 is broader than
the spectrum in Fig.~2).

\vspace{.18in}

Figure 13: Time-frequency distribution of the 4 millisec~time interval 
prior to the second ELM.
The large amplitude, 10 $\mu$sec~precursors burst is visible at 
$t =$ 12.734 msec.
The data has been filtered using a tenth order 
autoregressive filter to reduce the spectral range.
Prewhitening removes the fluctuations which are associated with
nonstationary activity of the central 1 MHz peak.
%just prior to the onset of the second ELM. 
The evolutionary spectrum 
is computed on 50-point segments with 50 \% overlap. We use
two Slepian tapers with a bandwidth of 100 kHz followed by a kernel smoother
with a kernel half-width of 100 kHz .

\vspace{.18in}

Figure 14: Time-frequency distribution of the prewhitened data
for the 4 millisec~time interval just prior to first ELM. 
The large amplitude, 10 $\mu$sec~precursors bursts are visible at 
$t =$  5.186 and  5.56 msec.
Only the first burst is visible on the AR
residual plot of Fig.~3b.

\vspace{.18in}

Figure 15a: Integrated energy in the frequency band $[300 -700]$ kHz
versus time for the first two ELMs.
The precursor bursts are clearly visible in the $300 -700$ kHz band.

\vspace{.15in}

Figure 15b: Integrated energy in two frequency band $[900 -1100]$ kHz
versus time.
The numerous bursts in the IF~frequency range occur with no clear pattern,
and thus are ill-suited to forecast ELM activity.  
% are barely visible in the electron drift wave frequency band.

\vspace{.15in}

Figure 15c: Integrated energy in the frequency band $[1300 -1700]$ kHz
versus time. 
The precursor bursts are barely visible in the electron drift wave
frequency band.
%Figure 12 shows quite clearly that the precursor bursts are localized
%in the $300 -700$ kHz band  which corresponds to ion drift fluctuations.
%4 millisec~time interval just prior to ? th ELM. 
\end{document}